%% file: LesHouchesmh.tex
\documentclass[12pt]{article}

\usepackage{epsfig,psfrag}

\catcode`@=11
\def\citer{\@ifnextchar
[{\@tempswatrue\@citexr}{\@tempswafalse\@citexr[]}}

\def\@citexr[#1]#2{\if@filesw\immediate\write\@auxout{\string\citation{#2}}\fi
  \def\@citea{}\@cite{\@for\@citeb:=#2\do
    {\@citea\def\@citea{--\penalty\@m}\@ifundefined
       {b@\@citeb}{{\bf ?}\@warning
       {Citation `\@citeb' on page \thepage \space undefined}}%
\hbox{\csname b@\@citeb\endcsname}}}{#1}}
\catcode`@=12

\newcommand{\beq}{\begin{eqnarray}}
\newcommand{\eeq}{\end{eqnarray}}
\newcommand{\bq}{\begin{equation}}
\newcommand{\eq}{\end{equation}}
\newcommand{\be}{\begin{equation}}
\newcommand{\ee}{\end{equation}}

\begin{document}

%%%%%%%%%%%%%%%%%%%%%%%%%%%%%%%%%%%%%%%%%%%%%%%%%%%%%%%%%%%%%%%%%%%%%%

%%%%%%%%%%%%%%%%%%%%%%%%%%%%%% NEW SHORT CUTS %%%%%%%%%%%%%%%%%%%%%%%%

\input paperdef 

\input{psfragdef}

%%%%%%%%%%%%%%%%%%%%%%%%%%%%%% TITLE PAGE %%%%%%%%%%%%%%%%%%%%%%%%%%%%

\thispagestyle{empty}
\setcounter{page}{0}
\def\thefootnote{\fnsymbol{footnote}}

\begin{flushright}
BNL--HET--01/38\\
DCPT/02/22\\
IPPP/02/10\\
KA--TP--2--2002\\
hep-ph/0202166 \\
\end{flushright}

%\vspace{1cm}

\begin{center}

{\large\sc {\bf  FeynHiggs1.2: Hybrid \msbarbf /on-shell Renormalization}}

\vspace*{0.4cm} 

{\large\sc {\bf for the $\cp$-even Higgs Boson Sector in the MSSM}}%
\footnote{Contribution to the workshop ``Physics at TeV Colliders'', 
 Les Houches, France, May 2001}

\vspace{1cm}

{\sc M. Frank$^{\,1}$%
\footnote{
email: Markus.Frank@physik.uni-karlsruhe.de
}%
, S. Heinemeyer$^{\,2}$%
\footnote{
email: Sven.Heinemeyer@physik.uni-muenchen.de 
%(BNL does not provide permanent email addresses)
}%
, W. Hollik$^{\,1, 3}$%
\footnote{
email: Wolfgang.Hollik@physik.uni-karlsruhe.de
}%
~and G. Weiglein$^{\,4}$%
\footnote{
email: Georg.Weiglein@durham.ac.uk
}%
}

\vspace*{1cm}

$^1$ Institut f\"ur Theoretische Physik, Universit\"at Karlsruhe, \\
D--76128 Karlsruhe, Germany

\vspace*{0.4cm}

$^2$ HET, Physics Department, Brookhaven Natl.\ Lab., Upton, NY 
11973, USA

\vspace*{0.4cm}    

$^3$ Max-Planck-Institut f\"ur Physik, F\"ohringer Ring 6,
D--80805 M\"unchen, Germany

\vspace*{0.4cm}

$^4$ Institute for Particle Physics Phenomenology, University of Durham,\\
Durham DH1~3LE, UK

\end{center}

\vspace*{1cm}

\begin{abstract}

An updated version is presented of the Fortran code
\fh\ for the evaluation of the neutral $\cp$-even Higgs
sector masses and mixing angles. It
differs from the previous version 
by a modification of the  renormalization scheme 
concerning the treatment of subleading terms at the \onel\
level; the two-loop corrections, for which the leading contributions
of \order{\alt\als} and \order{\alt^2} are implemented,
are not affected by the modified renormalization prescription.
Besides stabilizing the Higgs mass calculations
and avoiding unphysically large threshold effects, the mass of the 
lightest MSSM Higgs boson, $\mh$, is increased by 1-2 GeV for most
parts of the MSSM parameter space.
 
\end{abstract}

\def\thefootnote{\arabic{footnote}}
\setcounter{footnote}{0}

\newpage

%%%%%%%%%%%%%%%%%%%%%%%%%%%%%% LH TITLE PAGE %%%%%%%%%%%%%%%%%%%%%%%%%

\begin{center}
{\large\sc {\bf FeynHiggs1.2: Hybrid \msbarbf /on-shell Renormalization
 for the \cp-even Higgs Boson Sector in the MSSM
}}

\vspace{0.5cm}

{\sc
M. Frank, S. Heinemeyer, W. Hollik, G. Weiglein
}
\end{center}

\begin{abstract}
An updated version is presented of the Fortran code
\fh\ for the evaluation of the neutral $\cp$-even Higgs
sector masses and mixing angles. It
differs from the previous version 
by a modification of the  renormalization scheme 
concerning the treatment of subleading terms at the \onel\
level; the two-loop corrections, for which the leading contributions
of \order{\alt\als} and \order{\alt^2} are implemented,
are not affected by the modified renormalization prescription.
Besides stabilizing the Higgs mass calculations
and avoiding unphysically large threshold effects, the mass of the 
lightest MSSM Higgs boson, $\mh$, is increased by 1-2 GeV for most
parts of the MSSM parameter space.
\end{abstract} 

%%%%%%%%%%%%%%%%%%%%%%%%%%%%%%%  MAIN TEXT  %%%%%%%%%%%%%%%%%%%%%%%%%%%%

\section{Introduction}

The search for the lightest Higgs boson in the Minimal Supersymmetric
Standard Model (MSSM) is one of the main goals at the
present and the next generation of colliders. Therefore the precise
knowledge of the dependence of the masses and mixing angles of the
Higgs sector of the MSSM on the relevant supersymmetric
parameters is of high importance.

In this note we present an updated version of the Fortran code
\fh~\cite{feynhiggs} that evaluates the neutral $\cp$-even Higgs
sector masses and mixing angles~\cite{mhiggsletter,mhiggslong}. It
differs from the previous version as presented in 
\citere{feynhiggs} 
by a modification of the  renormalization scheme 
concerning the treatment of subleading terms at the \onel\
level; the two-loop corrections, for which the leading contributions
of \order{\alt\als} and \order{\alt^2} are implemented,
are not affected.
In particular, an \msbar\ 
renormalization for $\tb$ and the field renormalization constants has
been used (where the \msbar\ quantities are evaluated 
at the scale $\mt$). The renormalization in the new version
of \fh\ does no longer involve the derivative 
of the $A$~boson self-energy and the $AZ$~mixing self-energy. This
leads to a more stable behavior around 
thresholds, e.g.\ at $\MA \approx 2\,\mt$, and avoids unphysically large
contributions in certain regions of the MSSM parameter space.
Thus, the new renormalization scheme stabilizes the prediction of the
masses and mixing angles in the $\cp$-even Higgs sector of the MSSM.

%%%%%%%%%%%%%%%%%%%%%%%%%%%%%%%%%%%%%%%%%%%%%%%%%%%%%%%%%%%%%%%%%%%%%%%%
%%%%%%%%%%%%%%%%%%%%%%%%%%%%%%%%%%%%%%%%%%%%%%%%%%%%%%%%%%%%%%%%%%%%%%%%

\section{Renormalization schemes}

The Higgs sector of the MSSM~\cite{hhg} consists of two neutral
$\cp$-even Higgs bosons, $h$ and $H$ ($\mh < \mH$), the
$\cp$-odd $A$~boson, and two charged Higgs bosons, $H^\pm$.
At the tree-level, $\mh$ and
$\mH$ can be evaluated in terms of the Standard Model (SM) gauge
couplings and two additional MSSM parameters, conventionally chosen as
$\MA$ and $\tb$, the ratio of the two vacuum expectation values 
($\tb = v_2/v_1$). 
Beyond lowest order, the Feynman-diagrammatic (FD) approach allows to
obtain in principle the most precise evaluation of the neutral
$\cp$-even Higgs boson sector, since in this way
the effect of different mass scales of the supersymmetric
particles and of 
the external momentum can consistently be included. 
The masses of the two $\cp$-even Higgs bosons are obtained in this
approach by
determining the poles of the $h-H$-propagator
matrix, which is equivalent to solving the equation
\BE
[ \; q^2 - m_{h,{\rm tree}}^2 + \hat\Sigma_{h}(q^2) \; ]
[ \; q^2 - m_{H,{\rm tree}}^2 + \hat\Sigma_{H}(q^2) \; ] -
[ \; \hat\Sigma_{hH}(q^2) \; ]^2 = 0 ,
\label{eq:propmatrix}
\EE
where $\hat\Sigma_s, s = h, H, hH$, denote the renormalized Higgs
boson self-energies. For the renormalization within the FD approach
usually the on-shell scheme is
applied~\cite{mhiggslong}. This means in particular that all the masses
in the FD result are the physical ones, i.e.\ they correspond to
physical observables. Since \refeq{eq:propmatrix} is solved iteratively,
the result for $m_h$ and $m_H$ contains a dependence on the field
renormalization constants of $h$ and $H$, which is 
formally of higher order. Accordingly, there is some freedom in choosing 
appropriate renormalization conditions for fixing the field
renormalization constants (this can also be interpreted as affecting the
renormalization of $\tb$). Different renormalization conditions have
been considered, e.g.\ ($\hSip$ denotes the derivative with respect to
the squared momentum):
\begin{enumerate}
\item
on-shell renormalization for $\hSi_Z, \hSi_A,
\hSip_A, \hSi_{AZ}$, and  
$\de v_1/v_1 = \de v_2/v_2$~\cite{mhiggs1lfull}
\item
on-shell renormalization for 
$\hSi_Z, \hSi_A, \hSi_{AZ}$, and 
$\de v_i = \de v_{i, {\rm div}}, i = 1,2$~\cite{mhiggs1lfullb}
\item
on-shell renormalization for $\hSi_Z, \hSi_A$~\cite{mhiggs1lfull},
\msbar\ renormalization for $\de Z_h, \de Z_H$, $\tb$~\cite{mhiggsrenorm}.
\end{enumerate}
The previous version of \fh\ is based on renormalization~1, involving
the derivative of the $A$~boson self-energy. The new version of \fh,
see {\tt www.feynhiggs.de}, is based on renormalization~3 (a detailed 
discussion can be found in \citere{mhiggsrenorm}).

%%%%%%%%%%%%%%%%%%%%%%%%%%%%%%%%%%%%%%%%%%%%%%%%%%%%%%%%%%%%%%%%%%%%%%%%
%%%%%%%%%%%%%%%%%%%%%%%%%%%%%%%%%%%%%%%%%%%%%%%%%%%%%%%%%%%%%%%%%%%%%%%%

\section{Numerical comparison}

In this section we numerically compare the output of 
the previous version (based on renormalization~1) and the
new version (based on renormalization~3) of \fh.
We also show results for the recently obtained
non-logarithmic \order{\alt^2} corrections~\cite{mhalphatsq,maulpaul}
that are also included in the new version of \fh.
The comparison is performed for the parameters of the three LEP 
benchmark scenarios~\cite{benchmark}. In this way, the effect of the
new renormalization and the non-logarithmic \order{\alt^2} corrections
on the analysis of the LEP Higgs-boson searches can easily be read off.

\begin{figure}[htb!]
\mbox{}\vspace{-2cm}
\begin{center}
\mbox{
\epsfig{figure=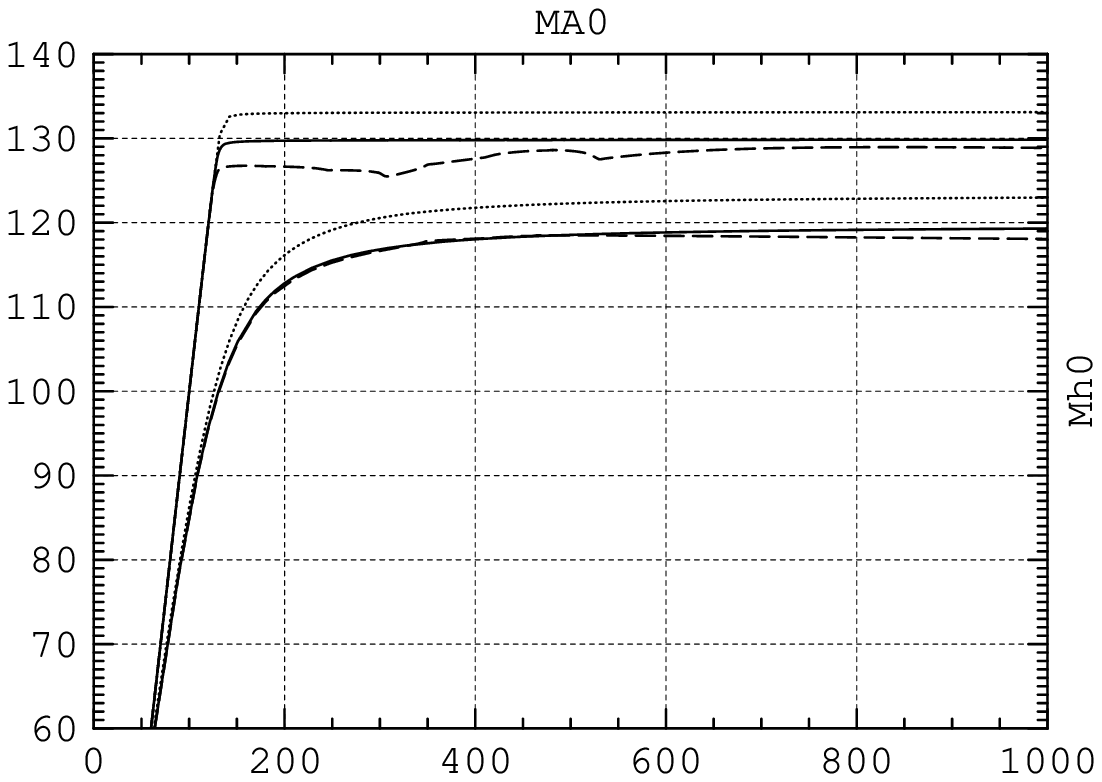,width=7cm,height=5.5cm} 
\hspace{1em}
\epsfig{figure=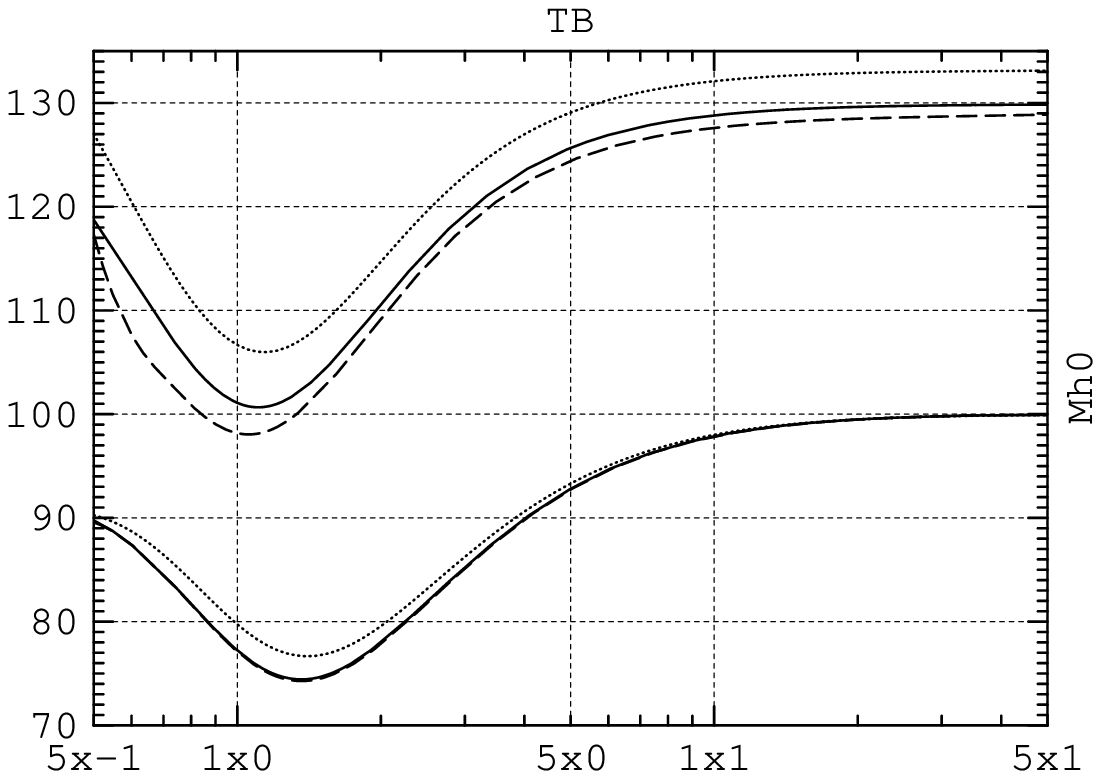,width=7cm,height=5.5cm}}
\end{center}
\vspace{-2.1em}
\caption[]{The new renormalization (3, solid) and the old scheme (1,
dashed) are compared in the $\mhmax$ scenario. The dotted line shows
the inclusion of the non-logarithmic \order{\alt^2} corrections. The 
lower curves are for 
$\tb = 3$ (left plot) or $\MA = 100 \gev$ (right). The upper curves are
for $\tb = 50$ (left) or $\MA = 1000 \gev$ (right).}
\label{fig:mhmax}
\end{figure}

\begin{figure}[htb!]
\begin{center}
\mbox{
\epsfig{figure=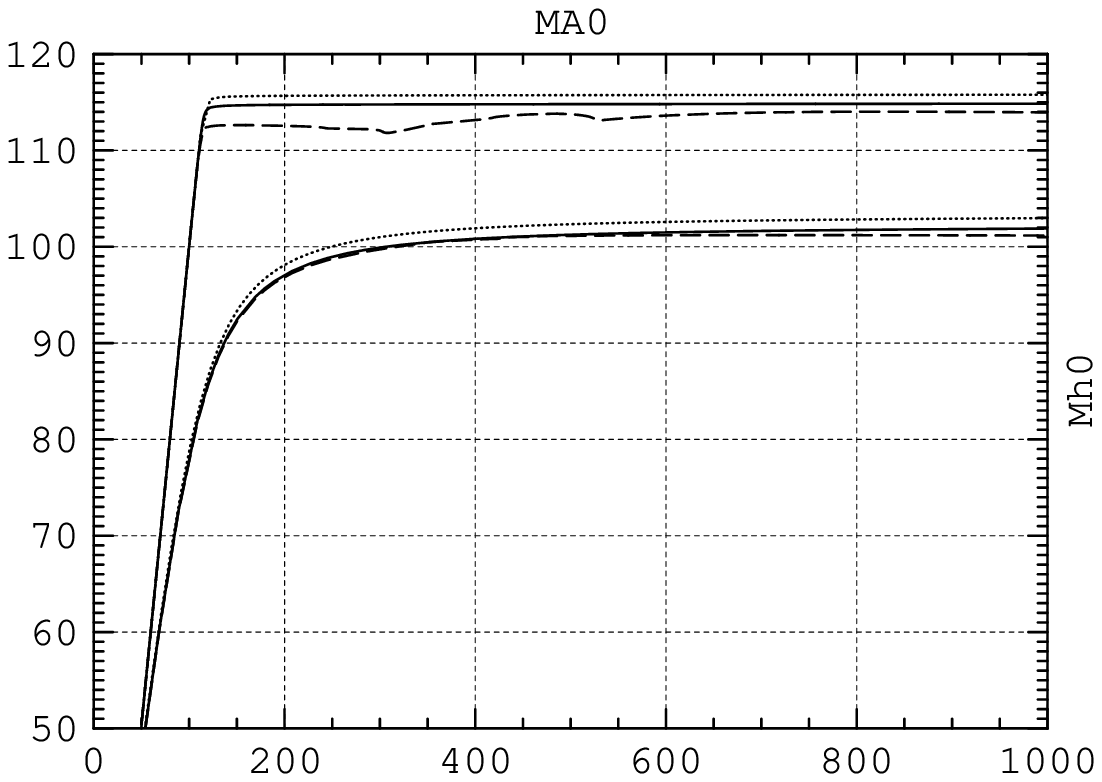,width=7cm,height=5.5cm} 
\hspace{1em}
\epsfig{figure=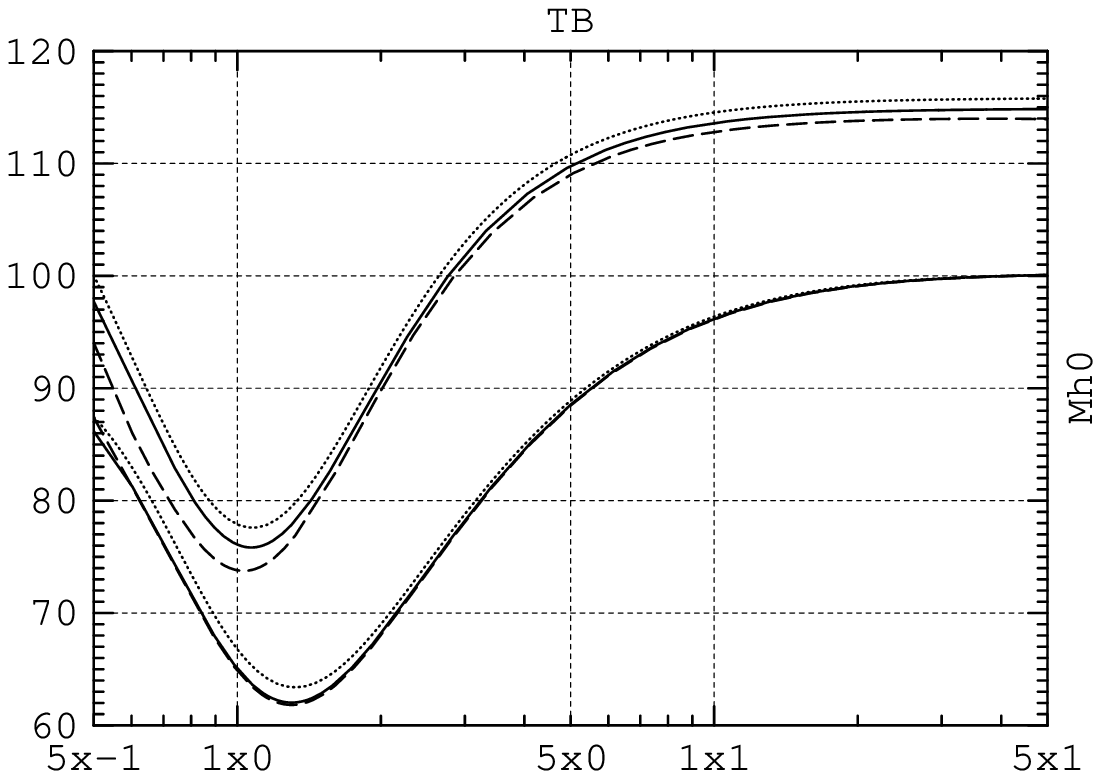,width=7cm,height=5.5cm}}
\end{center}
\vspace{-2.1em}
\caption[]{The new renormalization (3, solid) and the old scheme (1,
dashed) are compared in the no-mixing scenario. The dotted line shows
the inclusion of the non-logarithmic \order{\alt^2} corrections. The lower
curves are for 
$\tb = 3$ (left plot) or $\MA = 100 \gev$ (right). The upper curves are
for $\tb = 50$ (left) or $\MA = 1000 \gev$ (right).}
\label{fig:nomix}
\end{figure}

\begin{figure}[htb!]
\begin{center}
\mbox{
\epsfig{figure=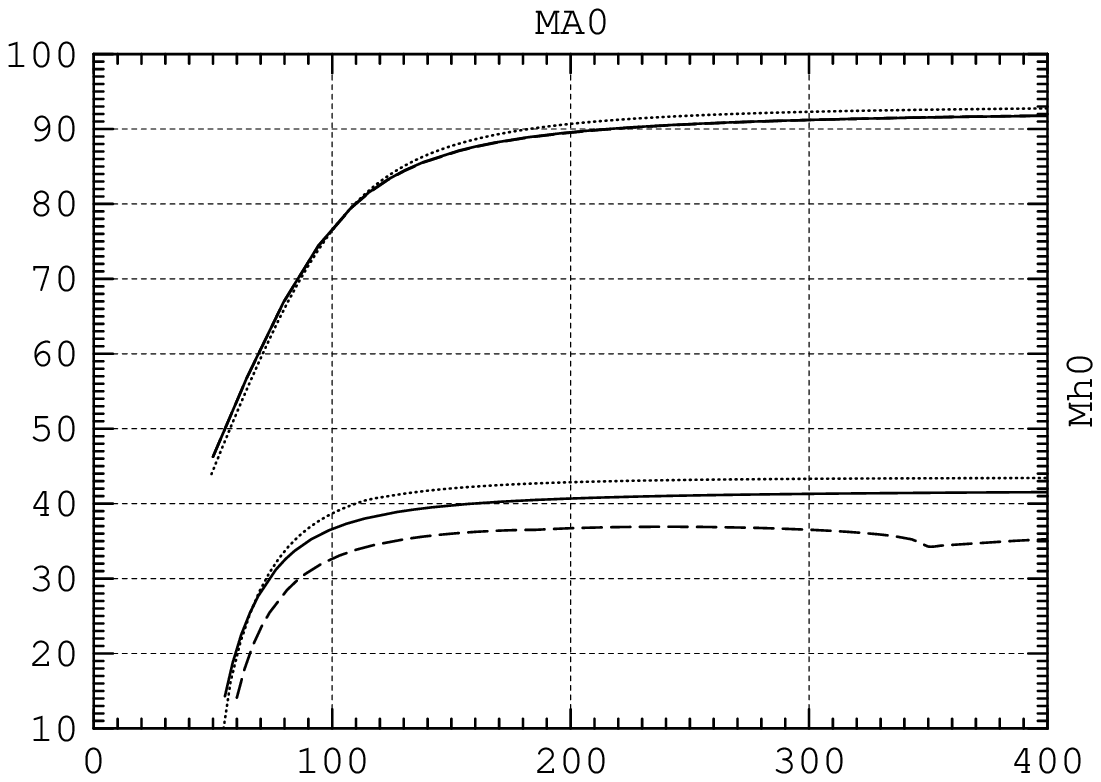,width=7cm,height=5.5cm} 
\hspace{1em}
\epsfig{figure=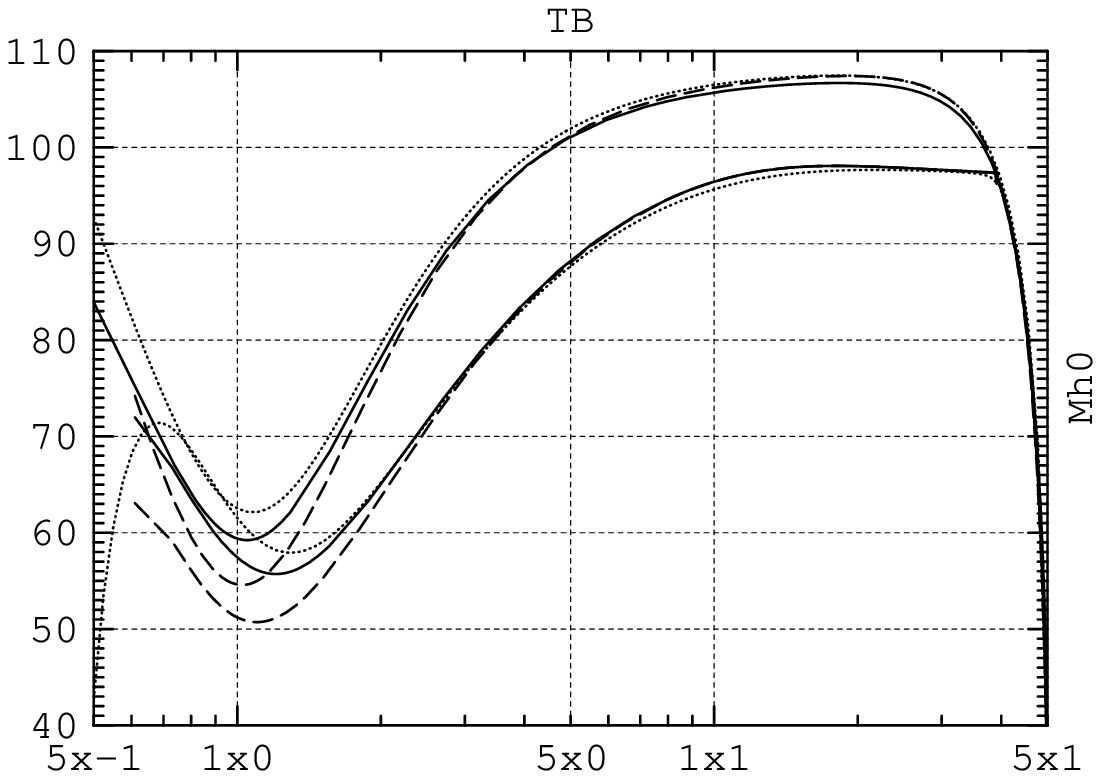,width=7cm,height=5.5cm}}
\end{center}
\vspace{-2.1em}
\caption[]{The new renormalization (3, solid) and the old scheme (1,
dashed) are compared in the large $\mu$ scenario. The dotted line shows
the inclusion of the non-logarithmic \order{\alt^2} corrections. The lower
curves are for 
$\tb = 50$ (left plot) or $\MA = 100 \gev$ (right). The upper curves are
for $\tb = 3$ (left) or $\MA = 400 \gev$ (right).}
\vspace{-0.6em}
\label{fig:largemu}
\vspace{-1em}
\end{figure}

In \reffis{fig:mhmax}--\ref{fig:largemu} we show the results in the
``$\mhmax$'', ``no-mixing'' and ``large $\mu$'' scenario as a function
of $\MA$ (left column) and of $\tb$ (right column) for two values of
$\tb$ ($\tb = 3, 50$) and $\MA$ ($\MA = 100, 1000 \gev$ for the $\mhmax$
and the no-mixing scenario, $\MA = 100, 400 \gev$ for the large~$\mu$
scenario), respectively. The solid lines correspond to the new 
result while the dashed lines show the old results. The dotted lines
correspond to the new result including the non-logarithmic
\order{\alt^2} contributions. Concerning the new renormalization
scheme, in the $\mhmax$
(\reffi{fig:mhmax}) and the no-mixing scenario (\reffi{fig:nomix}) the
new result is larger by $\approx$1--2~GeV for 
not too small $\MA$ and $\tb$. For small $\tb$ and large $\MA$ the
enhancement can be 
even larger. In the large $\mu$ scenario (\reffi{fig:largemu}) the largest
deviations appear for small $\tb$ for both large and small $\MA$. 
While the previous prescription for the field renormalization constants 
leads to
unphysically large threshold effects in some regions of the parameter
space, which arise from the $AZ$ mixing self-energy and the 
derivative of the $A$~boson
self-energy, no threshold kinks are visible for the result based on the
new renormalization.
The shift in $\mh$ of $\approx$1--2~GeV related to the modification of the
renormalization prescription lies in the range of the anticipated
theoretical uncertainty from unknown non-leading electroweak two-loop
corrections~\cite{mhiggsstatus}. 
The new \order{\alt^2} corrections can further increase $\mh$
by up to $\approx$3~GeV for large $\Stop$~mixing (a detailed analysis
will be presented elsewhere~\cite{mhalphatsqAnal}).

\smallskip
The new version of \fh\ can be obtained from {\tt www.feynhiggs.de} .

%%%%%%%%%%%%%%%%%%%%%%%%%%%%%%%%%%%%%%%%%%%%%%%%%%%%%%%%%%%%%%%%%%%%%%%%
%%%%%%%%%%%%%%%%%%%%%%%%%%%%%%%%%%%%%%%%%%%%%%%%%%%%%%%%%%%%%%%%%%%%%%%%

\vspace{-.5em}
\subsubsection*{Acknowledgements}

G.W.\ thanks the organizers of the Les Houches workshop for the
invitation and the pleasant and constructive atmosphere. 
%Only the opening hours of the bar could have been longer.
This work was supported in part by the European Community's Human
Potential Programme under contract HPRN-CT-2000-00149 Physics at Colliders.

%%%%%%%%%%%%%%%%%%%%%%%  REFERENCES  %%%%%%%%%%%%%%%%%%%%%%%

\bibliographystyle{plain}

%%%%%%%%%%%%%%%%%%%%% FIGURE %%%%%%%%%%%%%%%%%%%%%%%%%%%%%%%%%%%%%%%%%%
%%%%%%%%%%%%%%%%%%%%% FIGURE %%%%%%%%%%%%%%%%%%%%%%%%%%%%%%%%%%%%%%%%%%
%%%%%%%%%%%%%%%%%%%%% FIGURE %%%%%%%%%%%%%%%%%%%%%%%%%%%%%%%%%%%%%%%%%%
%%%%%%%%%%%%%%%%%%%%% FIGURE %%%%%%%%%%%%%%%%%%%%%%%%%%%%%%%%%%%%%%%%%%
%%%%%%%%%%%%%%%%%%%%% FIGURE %%%%%%%%%%%%%%%%%%%%%%%%%%%%%%%%%%%%%%%%%%
%%%%%%%%%%%%%%%%%%%%% FIGURE %%%%%%%%%%%%%%%%%%%%%%%%%%%%%%%%%%%%%%%%%%

\end{document}

%% file: psfragdef.tex
\newcommand{\psfragtextscale}{0.75}
\psfrag{Mh0}[][][\psfragtextscale]{$m_h [\mathrm{GeV}]$}
\psfrag{MA0}[][][\psfragtextscale]{$M_A [\mathrm{GeV}]$}
\psfrag{TB}[][][\psfragtextscale]{$\tan \beta$}
\psfrag{5x-1}[][][\psfragtextscale]{0.5}
\psfrag{1x0}[][][\psfragtextscale]{1}
\psfrag{5x0}[][][\psfragtextscale]{5}
\psfrag{1x1}[][][\psfragtextscale]{10}
\psfrag{5x1}[][][\psfragtextscale]{50}
\psfrag{0}[][][\psfragtextscale]{0}
\psfrag{1}[][][\psfragtextscale]{1}
\psfrag{2}[][][\psfragtextscale]{2}
\psfrag{3}[][][\psfragtextscale]{3}
\psfrag{4}[][][\psfragtextscale]{4}
\psfrag{5}[][][\psfragtextscale]{5}
\psfrag{10}[][][\psfragtextscale]{10}
\psfrag{20}[][][\psfragtextscale]{20}
\psfrag{30}[][][\psfragtextscale]{30}
\psfrag{40}[][][\psfragtextscale]{40}
\psfrag{50}[][][\psfragtextscale]{50}
\psfrag{60}[][][\psfragtextscale]{60}
\psfrag{70}[][][\psfragtextscale]{70}
\psfrag{80}[][][\psfragtextscale]{80}
\psfrag{90}[][][\psfragtextscale]{90}
\psfrag{100}[][][\psfragtextscale]{100}
\psfrag{110}[][][\psfragtextscale]{110}
\psfrag{120}[][][\psfragtextscale]{120}
\psfrag{130}[][][\psfragtextscale]{130}
\psfrag{140}[][][\psfragtextscale]{140}
\psfrag{150}[][][\psfragtextscale]{150}
\psfrag{160}[][][\psfragtextscale]{160}
\psfrag{170}[][][\psfragtextscale]{170}
\psfrag{180}[][][\psfragtextscale]{180}
\psfrag{190}[][][\psfragtextscale]{190}
\psfrag{200}[][][\psfragtextscale]{200}
\psfrag{300}[][][\psfragtextscale]{300}
\psfrag{400}[][][\psfragtextscale]{400}
\psfrag{500}[][][\psfragtextscale]{500}
\psfrag{600}[][][\psfragtextscale]{600}
\psfrag{700}[][][\psfragtextscale]{700}
\psfrag{800}[][][\psfragtextscale]{800}
\psfrag{900}[][][\psfragtextscale]{900}
\psfrag{1000}[][][\psfragtextscale]{1000}
\psfrag{1500}[][][\psfragtextscale]{1500}
\psfrag{-1}[][][\psfragtextscale]{-1}
\psfrag{-2}[][][\psfragtextscale]{-2}
\psfrag{-3}[][][\psfragtextscale]{-3}
\psfrag{-4}[][][\psfragtextscale]{-4}
\psfrag{-5}[][][\psfragtextscale]{-5}
\psfrag{-10}[][][\psfragtextscale]{-10}
\psfrag{-20}[][][\psfragtextscale]{-20}
\psfrag{-30}[][][\psfragtextscale]{-30}
\psfrag{-40}[][][\psfragtextscale]{-40}
\psfrag{-50}[][][\psfragtextscale]{-50}
\psfrag{-60}[][][\psfragtextscale]{-60}
\psfrag{-70}[][][\psfragtextscale]{-70}
\psfrag{-80}[][][\psfragtextscale]{-80}
\psfrag{-90}[][][\psfragtextscale]{-90}
\psfrag{-100}[][][\psfragtextscale]{-100}
\psfrag{-110}[][][\psfragtextscale]{-110}
\psfrag{-120}[][][\psfragtextscale]{-120}
\psfrag{-130}[][][\psfragtextscale]{-130}
\psfrag{-140}[][][\psfragtextscale]{-140}
\psfrag{-150}[][][\psfragtextscale]{-150}
\psfrag{-160}[][][\psfragtextscale]{-160}
\psfrag{-170}[][][\psfragtextscale]{-170}
\psfrag{-180}[][][\psfragtextscale]{-180}
\psfrag{-190}[][][\psfragtextscale]{-190}
\psfrag{-200}[][][\psfragtextscale]{-200}
\psfrag{-300}[][][\psfragtextscale]{-300}
\psfrag{-400}[][][\psfragtextscale]{-400}
\psfrag{-500}[][][\psfragtextscale]{-500}
\psfrag{-600}[][][\psfragtextscale]{-600}
\psfrag{-700}[][][\psfragtextscale]{-700}
\psfrag{-800}[][][\psfragtextscale]{-800}
\psfrag{-900}[][][\psfragtextscale]{-900}
\psfrag{-1000}[][][\psfragtextscale]{-1000}
\psfrag{-1500}[][][\psfragtextscale]{-1500}

%% file: LesHouchesmh.bbl
\begin{thebibliography}{99}


\bibitem{feynhiggs} S.~Heinemeyer, W.~Hollik and G.~Weiglein,
                    {\em Comp. Phys. Comm.} {\bf 124} (2000) 76;
%                    hep-ph/9812320;
                    %%CITATION = HEP-PH 9812320;%%
                    hep-ph/0002213;
                    %%CITATION = HEP-PH 0002213;%%
                    see {\tt www.feynhiggs.de} .

\bibitem{mhiggsletter} S.~Heinemeyer, W.~Hollik and G.~Weiglein, 
                       {\em Phys. Rev.} {\bf D 58} (1998) 091701;
%                       hep-ph/9803277; 
                       %%CITATION = HEP-PH 9803277;%%
                       hep-ph/9806250;
                       %%CITATION = HEP-PH 9806250;%%
                       {\em Phys. Lett.} {\bf B 440} (1998) 296.
%                       hep-ph/9807423.
                       %%CITATION = HEP-PH 9807423;%%

\bibitem{mhiggslong} S.~Heinemeyer, W.~Hollik and G.~Weiglein,
                     {Eur. Phys. Jour.} {\bf C 9} (1999) 343.
%                     hep-ph/9812472.
                     %%CITATION = HEP-PH 9812472;%%

\bibitem{hhg} J.~Gunion, H.~Haber, G.~Kane and S.~Dawson, 
              {\em The Higgs Hunter's Guide}, 
              Addison-Wesley, 1990.

\bibitem{mhiggs1lfull}  A.~Dabelstein, 
                        {\em Z. Phys.} {\bf C 67} (1995) 495;
%                        hep-ph/9409375;
                        %%CITATION = HEP-PH 9409375;%%
                        {\em Nucl. Phys.} {\bf B 456} (1995) 25.
%                        hep-ph/9503443.
                        %%CITATION = HEP-PH 9503443;%%

\bibitem{mhiggs1lfullb} P.~Chankowski, S.~Pokorski and J.~Rosiek,
                        {\em Nucl. Phys.} {\bf B 423} (1994) 437;\\
%                        hep-ph/9303309;
                        %%CITATION = HEP-PH 9303309;%%
                        S.~Heinemeyer, W.~Hollik, J.~Rosiek and G.~Weiglein, 
                        {\em Eur. Phys. Jour.} {\bf C 19}~(2001)~535.
%                        hep-ph/0102081.
                        %%CITATION = HEP-PH 9303309;%%

\bibitem{mhiggsrenorm} M.~Frank, S.~Heinemeyer, W.~Hollik and G.~Weiglein,
                       {\em in preparation}.

\bibitem{mhalphatsq} A. Brignole, G.~Degrassi, P.~Slavich and F.~Zwirner,
                     hep-ph/0112177.
                     %%CITATION = HEP-PH 0112177;%%

\bibitem{maulpaul} J.~Espinosa and R.-J.~Zhang, 
                   {\em Nucl. Phys.} {\bf B 586} (2000) 3.
%                   hep-ph/0003246.
                   %%CITATION = HEP-PH 0003246;%%

\bibitem{benchmark} M.~Carena, S.~Heinemeyer, C.~Wagner and G.~Weiglein,
                    hep-ph/9912223.
                    %%CITATION = HEP-PH 9912223;%%

\bibitem{mhiggsstatus} S.~Heinemeyer and G.~Weiglein,
                       hep-ph/0102117.
                       %%CITATION = HEP-PH 0102117;%%

\bibitem{mhalphatsqAnal} G.~Degrassi et al., 
                         {\em in preparation} .



\end{thebibliography}
